# High-speed metasurface modulator using critically coupled bimodal plasmonic resonance

*Jiaqi Zhang,*[1, b] *Yuji Kosugi,*[1, c] *Makoto Ogasawara,*[1, d] *Akira Otomo,*[2] *Toshiki Yamada,*[2] *Yoshiaki Nakano,*[1] *Takuo Tanemura*[1, a]

[1] School of Engineering, The University of Tokyo, 7-3-1 Hongo, Bunkyo-ku, Tokyo 113-8656, Japan.

[2] National Institute of Information and Communications Technology, 588-2 Iwaoka, Nishi-ku, Kobe 651-2492, Japan

[a] Author to whom correspondence should be addressed: tanemura@ee.t.u-tokyo.ac.jp

[b] Currently with Hisilicon Optoelectronics Ltd., China

[c] Currently with Sumitomo Electric Device Innovations, Inc., Japan

[d] Currently with Sumitomo Electric Industries Ltd., Japan

# ABSTRACT

Free-space electro-optic (EO) modulators operating at gigahertz and beyond are attractive for a wide range of emerging applications, including high-speed imaging, free-space optical communication, microwave photonics, and diffractive computing. Here we experimentally demonstrate a high-speed plasmonic metasurface EO modulator operating at a near-infrared wavelength range with a gigahertz modulation bandwidth. To achieve efficient intensity modulation of reflected light from an ultrathin metasurface layer, we utilize the bimodal plasmonic resonance inside a subwavelength metal-insulator-metal grating, which is precisely tuned to satisfy the critical coupling condition. As a result, perfect absorption of -27 dB (99.8%) and a high quality ($Q$) factor of 113 are obtained at a resonant wavelength of 1650 nm. By incorporating an EO polymer inside the grating, we achieve a modulation depth of up to 9.5 dB under an applied voltage of ±30 V. The 3-dB modulation bandwidth is confirmed to be 1.25 GHz, which is primarily limited by the undesired contact resistance. Owing to the high electrical conductivity of metallic gratings and a compact device structure with a minimal parasitic capacitance, the demonstrated device can potentially operate at several tens of gigahertz, which opens up exciting opportunities for ultrahigh-speed active metasurface devices in various applications.

**INTRODUCTION**

Metasurfaces, composed of dense arrays of artificially designed nanoscale optical resonators, offer unique means of manipulating lightwaves with ultrathin flat optics.[1,2] In addition to passive metasurface devices, there has recently been growing interest in developing actively reconfigurable metasurfaces.[3-6] In particular, electro-optic (EO) tuning of a plasmonic metasurface enables ultrahigh-speed intensity modulation of a normal-incident lightwave at subwavelength spatial resolution.[7-18] This will open up new opportunities for the metasurface devices in various applications, including high-speed imaging, free-space optical communication, microwave photonics, and diffractive computing.

A number of approaches have been demonstrated to date to realize plasmonic metasurfaces with active and high-speed EO materials, such as graphene,[7-9] highly doped semiconductors,[10] transparent conductive oxides,[11-16] thin-film lithium niobate,[17,18] and EO polymers.[19-23] Due to the short interaction lengths inside the thin metasurfaces, however, it has been difficult to simultaneously achieve high-speed operation and large modulation depth. As a common method to cope with the limited interaction length of an ultrathin metasurface, optical resonance is utilized. More specifically, a moderately high $Q$ factor in the order of few 100s is helpful in boosting the modulation efficiency, while maintaining the modulation bandwidth to be broad enough for many applications. In addition, to obtain a large modulation depth, it is essential to achieve perfect absorption based on the concept of critical coupling. While such scheme has been demonstrated in the mid-infrared range,[7,11,12] achieving critically coupled perfect absorption at the near-infrared wavelength range has been challenging due to the higher plasmonic absorption, leading to serious degradation of the $Q$ factor. As a result, previous experimental demonstrations of plasmonic metasurface EO modulators at the near-infrared range suffered from low $Q$ factor (<100) and/or small absorption dip (<10 dB), which are insufficient for achieving large modulation depth with minimal optical insertion loss.

In this paper, we experimentally demonstrate a high-speed (>GHz) electro-optic metasurface modulator operating at near-infrared wavelength (1650 nm) based on the critically coupled plasmonic resonator with the $Q$ factor exceeding 100. We employ EO polymer as the active material, which is embedded inside a 500-nm-thick metal-insulator-metal (MIM) structure. The EO polymer exhibits a large EO coefficient over 200 pm/V,[24-26] ultrahigh-speed modulation capability beyond 100 GHz,[27] simple and low-cost spin-coating-based fabrication that enables flexible device designs,[19-23,28-30] and high material reliability.[31,32] To reduce the radiation loss and achieve critical coupling without degrading the $Q$ factor, we exploit plasmonic bimodal resonance,[33] where the destructive interference of radiation from two MIM modes effectively suppresses the coupling to the radiative waves and boosts the $Q$ factor to 113. As a result, critical coupling condition is achieved with nearly perfect (less than -27 dB) absorption at the resonance wavelength. The intensity of the reflected light is modulated with an extinction ratio of 9.5 dB at ±30 V. The 3-dB modulation bandwidth of 1.25 GHz is obtained experimentally.

**DEVICE CONCEPT AND FABRICATION**

Figure 1(a) shows a schematic view of the high-speed metasurface modulator based on the plasmonic bimodal resonance. A 500-nm-thick EO polymer layer is sandwiched between a subwavelength Au grating and the bottom Au layer, which act as compact plasmonic resonators. By judiciously designing the grating parameters, we can permit only two transverse-magnetic (TM) MIM modes to exist inside the resonator.[33] When $x$-polarized light is input to the metasurface at a normal incident angle and when these two modes satisfy the resonant conditions, the light is strongly confined inside the EO polymer and eventually absorbed by the Au due to the inherent plasmonic loss. By applying external electric field to the EO polymer, the resonant wavelength is tuned so that the reflected light is intensity-modulated.

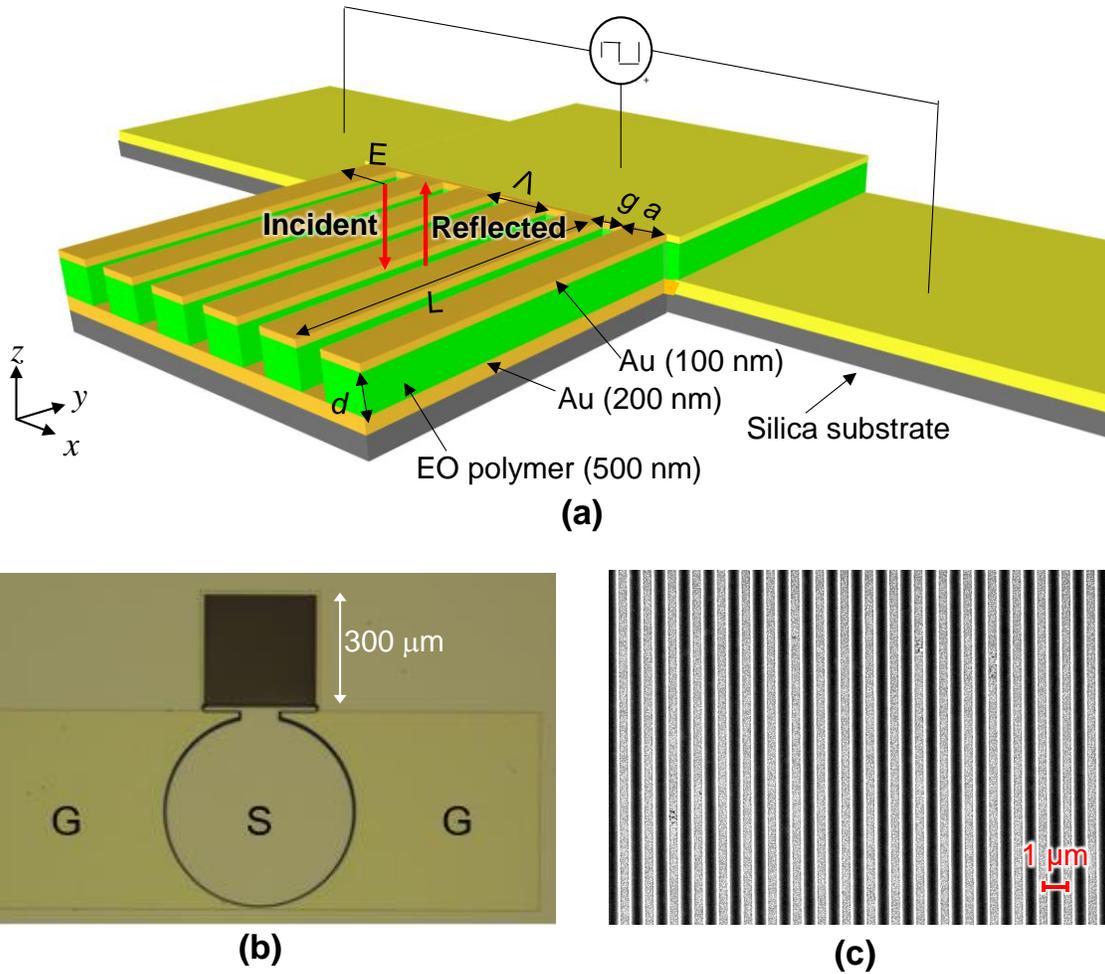

**FIG. 1.** Structure of the demonstrated metasurface modulator. (**a**) Schematic view. (**b**) Microscopic top view of the fabricated device. (**c**) Top SEM image of the grating with $\Lambda = 1080$ nm.

We should note that in this scheme, the metallic loss, which is often regarded as an unwanted feature that limits the performance of plasmonic devices, is utilized to obtain absorptive modulation. Since the off-resonant light is reflected at the top Au layer without experiencing a large metallic loss, the insertion loss of this modulator is inherently small, despite that it uses the plasmonic effect for absorptive modulation. Owing to the high electrical conductivity of Au ($4.2\times10^7$ S·m$^{-1}$) and relatively thick EO polymer (500 nm) with a small RF permittivity ($\varepsilon_{EO} = 2.7$), the resistance and the capacitance of the metasurface region itself are estimated to be as small as 0.5 Ω and 2.2 pF, respectively, for a 300 μm × 300 μm device [i.e., $L$ = 300 μm in Fig. 1(a)]. In this case, the inherent RC bandwidth of this device can be as high as 70 GHz. We assume that the resonator has a moderate $Q$ factor in the order of few 100s, so that this RC bandwidth would be the dominant bandwidth-limiting factor.

To obtain efficient modulation with a large extinction value, it is essential to obtain moderately high $Q$ factor (~100s) and satisfy the critical coupling condition, which enables complete light absorption. The critical coupling is achieved when the photon decay rate through the external radiation and that through the intrinsic plasmonic loss are precisely balanced.[7,12] However, in the conventional schemes based on a single resonant mode,[19,20] it is difficult to reduce the external radiation while keeping the plasmonic loss low, resulting in a degraded $Q$ factor. Here, we solve this problem by using the bimodal plasmonic resonance.[33] The effect of bimodal plasmonic resonance is described in Fig. 2(b), where the simulated reflectance is plotted as a function of wavelength λ and the grating period Λ with the gap $g$ fixed to 450 nm. When the two dispersion curves for the TM$_{0,2}$ and TM$_{1,2}$ resonant modes approach the anti-crossing point, these two modes interfere destructively to reduce the external radiation substantially. As a result, we can obtain a sharp and deep absorption dip in the reflectance spectrum at λ ~ 1650 nm. We should note that such anti-scattering behaviour is a general feature of strongly coupled multiple resonances[34,35] and has been observed in various systems, such as coupled ring resonators,[36] optical anapole resonators,[37] and high-contrast gratings.[28,29,38]

A set of devices with $L = 300$ μm and different values of the grating period $\Lambda$, ranging from 960 nm to 1100 nm, were fabricated on a silica substrate. First, a 200-nm-thick bottom Au layer was RF-sputtered with a thin Cr layer inserted between the silica substrate and Au to increase adhesion. To reduce the parasitic capacitance at the electrode pad and minimize the risk of electrical short circuit, the bottom Au layer underneath the top signal electrode pad was removed by a wet Au etchant. Then, a 10-nm-thick $Al_2O_3$ layer was deposited on the Au layer to avoid the device breakdown and the diffusion of Au during the poling process. Next, side-chain EO polymer[39] with a glass transition temperature ($T_g$) of 132.6 °C was deposited by spin-coating process. The thickness of the polymer was controlled by carefully adjusting the spin-coating conditions. After the post-annealing process, 100-nm-thick top Au layer was sputtered on the EO polymer surface. Subsequently, poling of the EO polymer was carried out by applying an electrical field of 80 V/μm between the top and bottom Au layers at 123 °C. In this condition, the EO coefficient $r_{33}$ was expected to be around 48 pm/V at 1550 nm wavelength. Finally, the subwavelength grating structure and top electrode pads were formed by electron-beam lithography (EBL) using ZEP520A resist, followed by the etching of Au and EO polymer layers using Ar-based inductively-coupled-plasma reactive-ion etching (ICP-RIE) and $O_2$-based RIE, respectively. Figures 1(b) and 1(c) show the top photograph and the scanning electron microscope (SEM) image of the fabricated device with $\Lambda = 1080$ nm.

**EXPERIMENT**

A linearly polarized (the electric field perpendicular to the grating) broadband incoherent beam was incident onto the device at normal incidence and the reflected spectrum was observed by an optical spectrum analyzer. The reflectance was normalized by the measured reflected power at an un-patterned region, which acted as a nearly ideal Au mirror. During the measurement, no degradation of the polymer layer was observed.

Figure 2(a) shows the reflectance spectra measured under the unbiased condition for the eight fabricated devices with different values of $\Lambda$. We can see that the measured resonant curves agree well with the simulated results shown in Fig. 2(b). More importantly, the $Q$ factor increases as the two resonance curves approach the anti-crossing point (more experimental data are provided in Supplementary Material to support this finding as well as clear anti-crossing splitting behaviour). In Fig. 2(a), a maximum $Q$ factor of 201 is obtained when $\Lambda = 1040$ nm. By carefully tuning $\Lambda$ around this bimodal resonance regime, we find that the critical coupling condition is fulfilled when $\Lambda = 1080$ nm. The measured reflectance spectrum in this case is shown in Fig. 2(c). We can confirm a sharp and deep resonance dip with a minimum reflectance of -27 dB (99.8% absorption) and a $Q$ factor of 113 at a wavelength of 1650 nm.

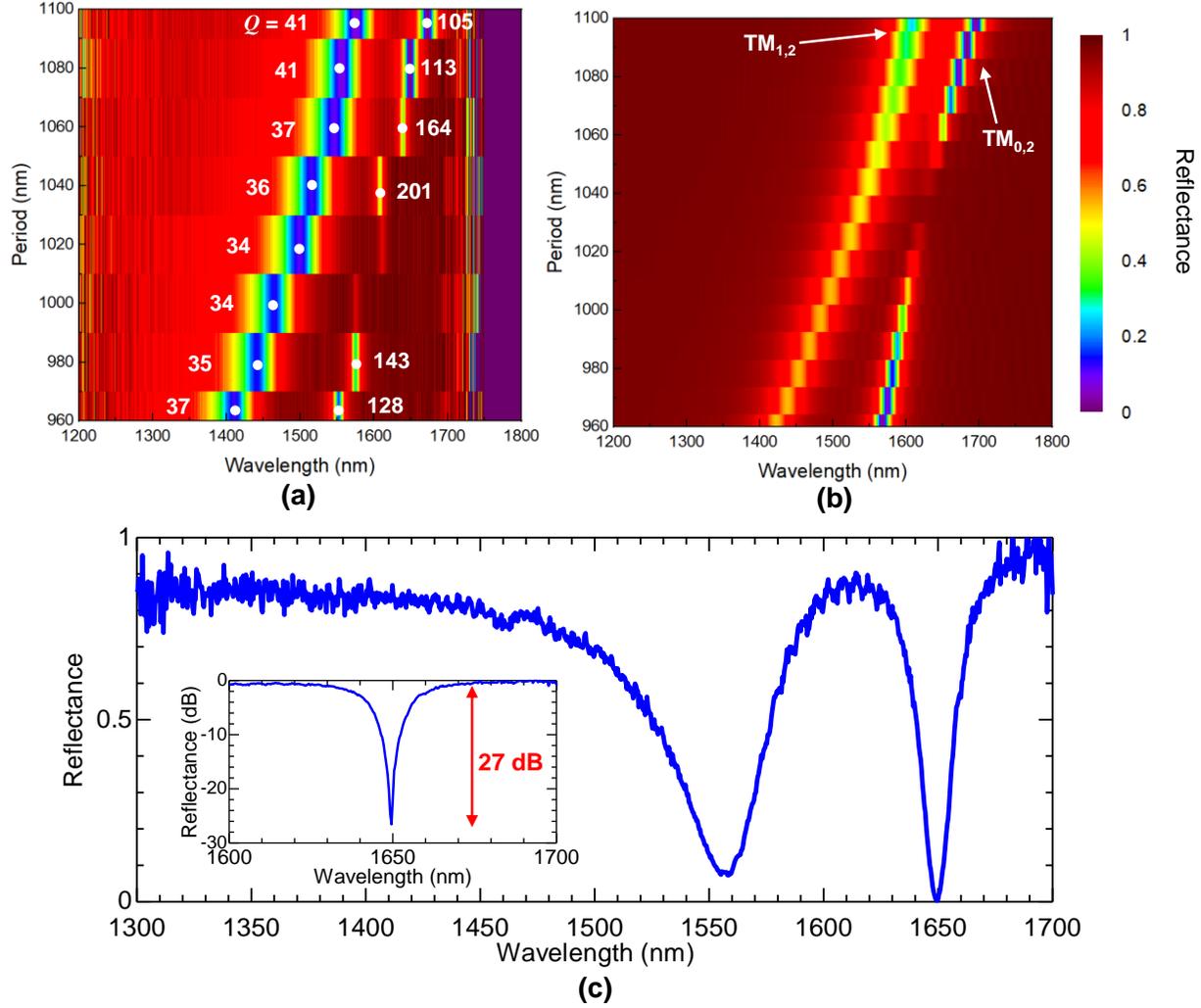

**FIG. 2.** Measured (**a**) and simulated (**b**) reflectance as a function of the grating period $\Lambda$ and the wavelength $\lambda$. (**c**) Measured reflectance spectrum for $\Lambda = 1080$ nm. The magnified spectrum near the resonance plotted on a log (dB) scale is shown in the inset.

Figure 3(a) shows the measured reflectance spectrum when a DC voltage is applied between the top and bottom Au layers for the device with $\Lambda = 1080$ nm. As the voltage increases (corresponding to a decrease in $\Delta n_z$), we see a clear blue shift in the resonant wavelength. From this result, the extinction ratio and the insertion loss of the modulator are derived as shown in Fig. 3(b). Once again, the insertion loss is defined with respect to the measured reflected power in the un-patterned region, which acted as a nearly ideal Au mirror. At an applied voltage of ±30 V, the highest extinction ratio of 9.5 dB with an insertion loss of 10.8 dB is obtained at a wavelength of 1652 nm. The insertion loss can be reduced to 5 dB if we operate at 1655 nm at the cost of decreased extinction ratio to 3.4 dB.

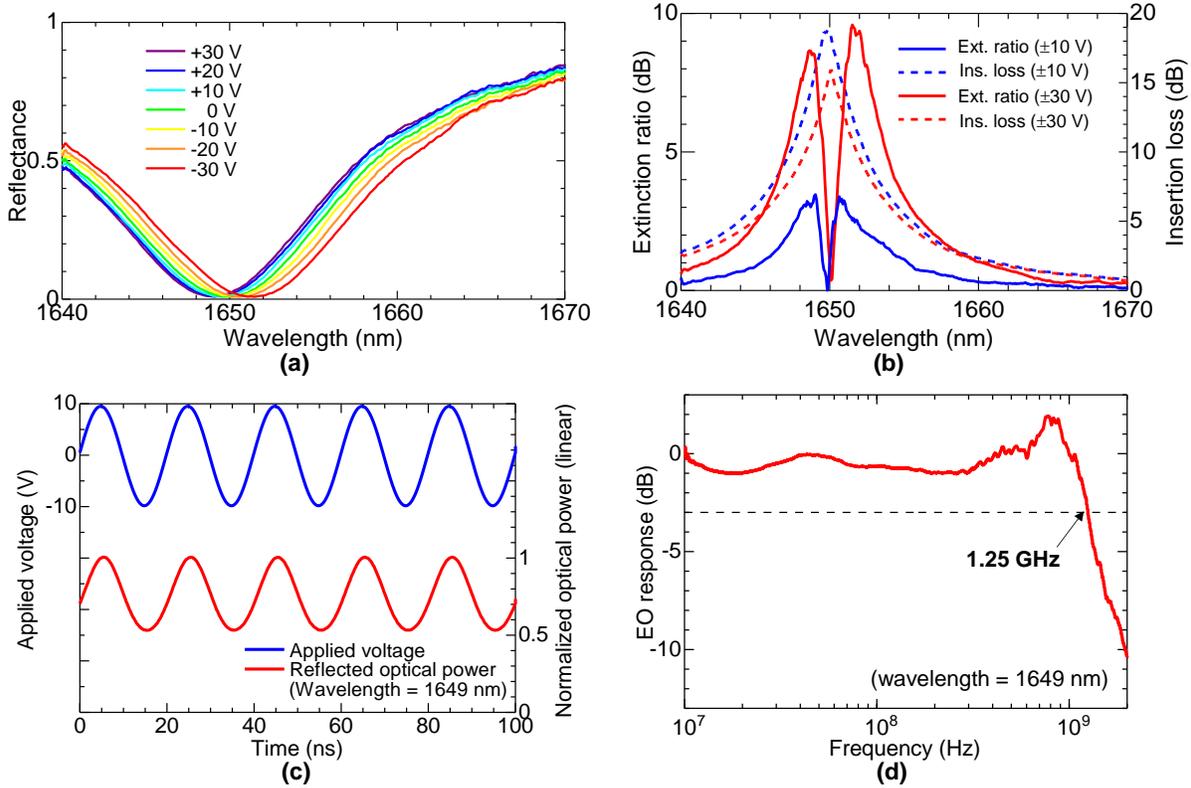

**FIG. 3.** Measured modulation characteristics for the device with $\Lambda = 1080$ nm. (**a**) Shift in optical reflection spectrum when a DC voltage from –30 V to +30 V is applied between the top and bottom Au layers. (**b**) Extinction ratio and insertion loss obtained at DC voltages of ±10 V and ±30 V. (**c**) Optical waveform at 1649 nm modulated by a 50-MHz 20-$V_{pp}$ sinusoidal signal. (**d**) EO frequency response of reflected light at 1649 nm.

By comparing the measured spectral shift in Fig. 3(a) with simulation, we deduce the actual $r_{33}$ of EO polymer in the fabricated devices to be around 14 pm/V. This corresponds to approximately seven-times improvement over our previous work,[20] but is still smaller than the value (48 pm/V) expected for the polymer material used in this work. The degradation in $r_{33}$ is attributed to the EBL and RIE processes after poling the EO polymer, which may have caused thermal relaxation of the poled chromophore to some extent. The use of more efficient EO polymer[24-26] with $r_{33}$ exceeding 100 pm/V should result in an increase in modulation efficiency by at least one order.

Finally, dynamic modulation properties are characterized at a wavelength of 1649 nm. Figure 3(c) shows the observed waveform when a 50-MHz 20-$V_{pp}$ sinusoidal electrical signal is applied. The extinction ratio of the modulated signal is around 48% (2.8 dB), which is consistent with the static value in Fig. 3(b). Figure 3(d) shows the small-signal frequency response. The 3-dB bandwidth is confirmed to be 1.25 GHz.

To clarify the limiting factor of the modulation bandwidth, the parasitic capacitance and the electrical resistance of the fabricated device were characterized. The measured capacitance between the ground and signal electrodes were measured to be around 3.2 pF, which is comparable to the theoretically estimated value of 2.1 pF. On the other hand, the electrical resistance was around 25 Ω, which is at least one-order larger than that calculated from the conductivity of Au grating. As a result, the RC bandwidth is derived to be around 2 GHz, which agrees reasonably with the measured result shown in Fig. 3(d). We therefore deduce that the modulation bandwidth of the current device is limited by the electrical contact resistance, rather than the response time of the EO polymer or the capacitance of the MIM structure itself. Higer-speed operation should thus be expected by reducing the contact resistance.

**CONCLUSIONS**

In summary, we have experimentally demonstrated high-speed modulation of plasmonic metasurface perfect absorber with EO polymer by exploiting the critically coupled bimodal resonance. By judiciously adjusting the grating period and measuring the reflectance spectrum of all samples, we obtained clear experimental evidence of a *Q* factor increase in the vicinity of the plasmonic bimodal resonance point. Using the optimized device with a grating period of 1080 nm, we achieved nearly perfect light absorption of -27 dB with a *Q* factor of 113 at a wavelength of 1650 nm. We then demonstrated dynamic tuning of the metasurface with the intensity modulation depth of up to 9.5 dB under an external voltage of ±30 V. The 3-dB modulation bandwidth was 1.25 GHz, which was limited by the relatively large (~25 Ω) contact electrical resistance. Further improvements in the modulation efficiency and the modulation bandwidth reaching tens of gigahertz are expected by using an EO polymer with higher $r_{33}$ and improving the electrical contact. With the potential of integrating large-scale two-dimensional pixels on a compact chip, this device would be attractive for various applications, such as free-space optical communication, high-speed imaging, and computing.

## SUPPLEMENTARY MATERIAL

Supplementary material provides additional experimental and numerical results of passive device to demonstrate the anti-crossing splitting behavior and $Q$-factor improvement.

## ACKNOWLEDGMENT

This work was supported in part by the Japan Science and Technology Agency (JST) PRESTO. A part of the device fabrication was conducted at Takeda Sentanchi Supercleanroom, The University of Tokyo, supported by the "Nanotechnology Platform Program" of the Ministry of Education, Culture, Sports, Science and Technology (MEXT), Japan. The authors acknowledge Ya-Lun Ho and Jean-Jacques Delaunay for their support in device fabrications. T.T. acknowledges Taichiro Fukui and Hiroki Miyano for the fruitful discussions during the preparation of the manuscript.

## AUTHOR DECLARATIONS

**Conflict of Interest**

The authors have no conflicts to disclose.

**Author Contributions**

T.T. conceived the idea and supervised the project. J.Z. designed and fabricated the devices and performed simulations and experiments. Y.K. and M.O. supported the device fabrication. A.O. and T.Y. developed and provided the EO polymer. Y.N., and T.T. contributed to developing the device fabrication and characterization infrastructures. T.T. and J.Z. wrote the manuscript. All authors contributed to the analysis of the results and revising the manuscript.

## DATA AVAILABILITY

The data that support the findings of this study are available from the corresponding author upon reasonable request.

# Supplementary Material for

# High-speed metasurface modulator using critically coupled bimodal plasmonic resonance

Jiaqi Zhang,[†] Yuji Kosugi,[†] Makoto Ogasawara,[†] Akira Otomo,[‡] Toshiki Yamada,[‡] Yoshiaki Nakano,[†] Takuo Tanemura[†, *]

[†] *School of Engineering, The University of Tokyo, 7-3-1 Hongo, Bunkyo-ku, Tokyo 113-8656, Japan.*

[‡] *National Institute of Information and Communications Technology, 588-2 Iwaoka, Nishi-ku, Kobe 651-2492, Japan*

*\* Corresponding author. E-mail: tanemura@ee.t.u-tokyo.ac.jp*

## 1. Detailed experimental investigation of *Q*-factor improvement

To investigate the behavior of *Q*-factor increase at the anti-crossing point of bimodal resonance in detail, we fabricated a passive device, containing 30 different designs with a broader range of the grating period Λ from 600 nm to 1180 nm. Figure S1 shows a schematic and scanning electron microscope (SEM) image of the fabricated device, as well as measured and simulated reflectance spectra. Note that we can observe some residue of etched Au piled up at the edges of the top surfaces of the Au bars (Fig. S1b). To consider this effect, the structure shown in Fig. S1a is assumed in the simulation shown in Fig. S1d. We can confirm rapid increase in *Q* factor as well as perfect absorption at the vicinity of the anti-crossing point.

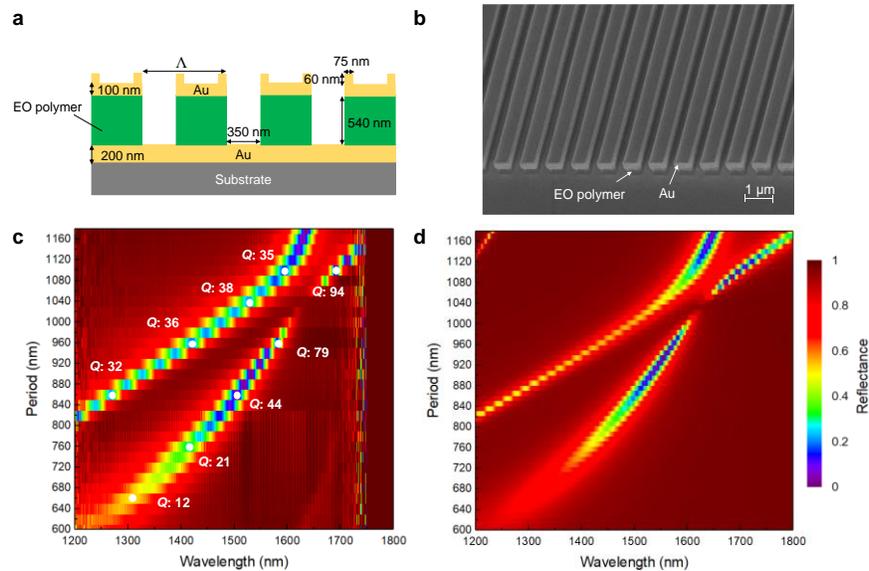

**Fig. S1:** Static reflectance measurements of an additionally fabricated sample. (**a**) The cross-section of the fabricated devices, having a broader range of the grating period Λ from 600 nm to 1180 nm. (**b**) SEM image of a fabricated structure with Λ = 1000 nm. (**c**) Measured and (**d**) simulated reflectance spectra for the structures with various Λ.

## 2. Numerical simulation and analysis of *Q* factor

To grasp the mechanism behind the improvement in *Q* factor and evidence of critical coupling, we numerically extract $1/\tau_{loss}$ and $1/\tau_{ext}$ of the $TM_{0,2}$ resonant mode in the vicinity of the anti-crossing point shown in Fig. S1. The reflection spectrum and *Q* factors at resonant wavelengths were calculated numerically by using a two-dimensional (2D) finite-difference time-domain (FDTD) solver from Lumerical. Periodic boundary conditions were employed in the *x* direction, while absorptive conditions with perfect matched layers (PMLs) were applied at *z* boundaries. The dielectric permittivity of the EO polymer was represented as a tensor, where the refractive indices ($n_x$, $n_y$, and $n_z$) were set to be 1.65 in the un-modulated condition. For simulating the case with a modulation field applied in the *z* direction (the same direction as the poling field), the refractive index tensor was modulated following the relation $\Delta n_x = \Delta n_y = 1/3 \Delta n_z$. This relation comes from the fact that $r_{13}$ and $r_{23}$ are approximately 1/3 of $r_{33}$, which is a valid assumption for a realistic poling voltage [1, 2]. The complex permittivity of Au was modelled by fitting the Drude-Lorentzian poles to the experimental data [3].

The *Q* factor of a plasmonic resonator is generally expressed as [4]

$$\frac{1}{Q} = \frac{1}{Q_{loss}} + \frac{1}{Q_{ext}} = \frac{\lambda}{\pi c \tau} = \frac{\lambda}{\pi c \tau_{loss}} + \frac{\lambda}{\pi c \tau_{ext}} \ ,$$

where $1/\tau_{loss}$ and $1/\tau_{ext}$ represent the photon decay rates due to the plasmonic loss and the external radiation, respectively, and $1/\tau$ is their sum, representing the total decay rate. $Q_{loss}$ and $Q_{ext}$ denote the *Q* factor, corresponding to respective decay rates, $\lambda$ is the resonant wavelength and *c* is the speed of light. In order to numerically extract $Q_{loss}$ and $Q_{ext}$ from the total *Q*, we represent the complex permittivity of Au $\varepsilon_{Au}$ as

$$\varepsilon_{Au} = \varepsilon_r + \xi \cdot i \varepsilon_i \ ,$$

where $\varepsilon_r$ and $\varepsilon_i$ are the real and imaginary components of $\varepsilon_{Au}$. A dimensionless coefficient $\xi$ ($0 \leq \xi \leq 1$) is used to artificially reduce the magnitude of plasmonic loss; when $\xi = 0$, it corresponds to an ideal Au without a plasmonic loss, so that $1/Q = 1/Q_{ext}$. Assuming that $1/Q_{loss}$ increases linearly with $\xi$ (which is a valid assumption if the mode profile does not change largely with $\xi$), both $1/Q_{loss}$ and $1/Q_{ext}$ can be derived by simulating the resonant spectrum for various values of $\xi$.

As example cases, Fig. S2a shows the simulated $1/Q$ as a function of $\xi$ for $\Lambda$ = 1000, 1033 and 1150 nm for the grating structure shown in Fig. S1. From the linear fit to each set of data, we can derive $Q_{ext}$ and $Q_{loss}$, which are then used to plot $1/\tau_{loss}$ and $1/\tau_{ext}$ in Fig. S2b. Their sum $1/\tau$ (= $1/\tau_{ext} + 1/\tau_{loss}$) is also plotted as a function of the grating period $\Lambda$. When $\Lambda$ is reduced from 1150 nm to 990 nm, $1/\tau_{ext}$ decreases by more than two orders of magnitude, from $1.8 \times 10^{13}$ 1/sec to $8.3 \times 10^{10}$ 1/sec. This reduction is due to the destructive interference of the radiation from the $TM_{0,2}$ and $TM_{1,2}$ modes, which strictly suppresses the coupling to the external wave. In contrast to $1/\tau_{ext}$, $1/\tau_{loss}$ remains almost independent of $\Lambda$, since the plasmonic loss is determined simply by the thickness of the EO polymer and the material properties of Au. When $\Lambda$ = 1033 nm, $1/\tau_{ext} = 1/\tau_{loss}$, so that the critical coupling condition is satisfied. As a result, a sharp absorption dip with nearly perfect absorption can be obtained, similar to the one observed in Fig. 2(c) in the main text.

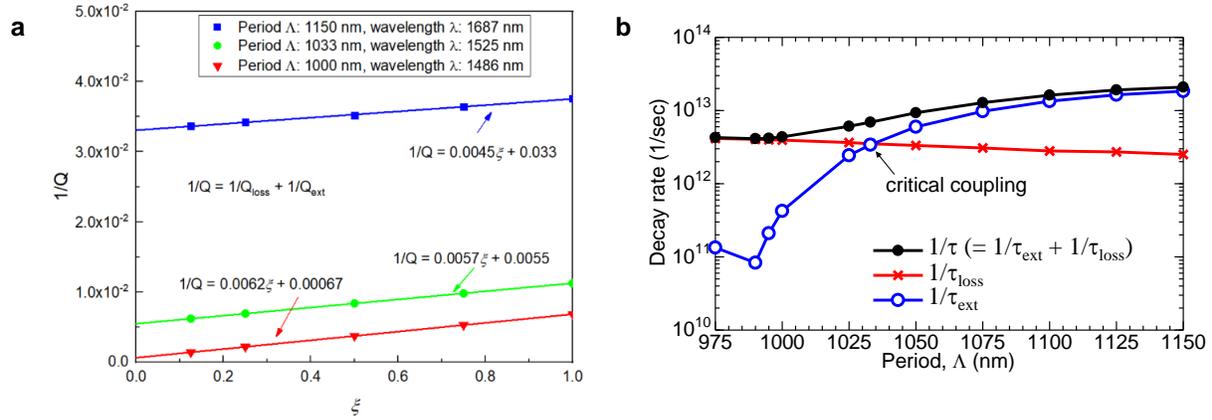

**Fig. S2:** Numerical analysis of the $Q$ factor and the photon decay rate of $TM_{0,2}$ resonant mode. (**a**) Simulated $1/Q$ as a function of $\xi$ for $\Lambda$ = 1150, 1033 nm and 1000 nm. (**b**) Decay rate of the $TM_{0,2}$ resonant mode for different values of $\Lambda$.